\documentclass[games,article,accept,moreauthors,12pt,a4paper]{mdpi} 
 \lastpage{x} \doinum{10.3390/------}
\pubvolume{xx} \pubyear{2012}
\history{Received: 2 February 2012; in revised form: 27 February 2012 / Accepted: 29 February 2012  / \\
Published: xx}

\usepackage{hyphenat}
\newcommand{\ts}[1]{\textsuperscript{#1}}
\usepackage{booktabs,enumerate}

\usepackage{amssymb,amsmath}

\Title{Coordination, Differentiation and Fairness in a Population of
Cooperating Agents}

\Author{Anne-Ly Do \ts{1,}*, Lars Rudolf \ts{2} and Thilo Gross
\ts{2}}

\address{%
\ts{1} Max Planck Institute for the Physics of Complex Systems, N\"othnitzer Str. 38, \linebreak  Dresden 01187, Germany\\
\ts{2} Department of Engineering Mathematics, University of Bristol,
Merchant Venturers Building, Bristol~BS8~1TR, UK; E-Mails: rudolflars@gmail.com (L.R.); thilo2gross@gmail.com (T.G.)
}

\corres{E-Mail:~ly@pks.mpg.de; Tel.:~+49-351-871-1119;
Fax:~+49-351-871-1299.}

\abstract{In a recent paper, we analyzed the self-assembly of a
complex cooperation network. The network was shown to approach a
state  where every agent invests the same amount of resources.
Nevertheless, highly-connected agents arise that extract
extraordinarily high payoffs while contributing comparably little to
any of their cooperations. Here, we investigate a variant of the
model, in which highly-connected agents have access to additional
resources. We study analytically and numerically whether these
resources are invested in existing collaborations, leading to a
fairer load distribution, or in establishing new collaborations,
leading to an even less fair distribution of loads and payoffs.}


\begin{document}

\section{Introduction}
Cooperation shapes our lives on many different
scales~\cite{AxelrodHamilton,DoebeliHauert}: Humans cooperate in
communities, companies, ethnicities and nations~\cite{Gulati,
Raustiala,Acharya}. Thereby, the collaborative behavior of the
individuals is strongly influenced by the embedding social
structure, the partners' social positions, and by social
norms~\cite{Gould,Willers,FehrFischbacher,Schmidt}. Conversely,
social structures, social positions, and social norms evolve in
response to the individuals' collaborative behavior~\cite{Macy,
Vos}.

The adaptive interplay between social structure and social behavior
has mostly been studied in a specific evolutionary context: Previous
models of human cooperation studied the influence of external
factors such as social
structure~\cite{NowakMay,HauertDoebeli,SantosPacheco,SantosSantos,Pacheco,Zschaler},
and social norms~\cite{BowlesGintis,OhtsukiHauert,HauertTraulsen} on
an individual's collaborative behavior. Here, we adopt the
complementary non-evolutionary perspective, asking for the
implications of an individual's behavior on the emergence of social
structures, social norms, and on the rise of leaders that hold
distinguished social positions.

In a recent paper, we considered a model which allows agents to
maintain different levels of cooperation with different self-chosen
partners and adapt them in response to their partners'
behavior~\cite{DoRudolfGross}. This revealed that a high degree of
social coordination can arise purely from the selective and adaptive
interaction of self-interested agents even if no social norm is
imposed externally: Although the agents possess little information,
the system approaches a state in which every agent makes the same
cooperative investment and every social interaction produces the
same benefit. We note that this coordination was not imposed
externally; different levels of investment evolved when the model
was run multiple times from effectively identical initial
conditions.

Despite the emergent coordination of investments, the final
configuration is generally not fair. Although we start the model
in an initially symmetric configuration which gave neither agent an
advantage, some agents manage to secure positions of high
centrality, where they interact with many other agents. In these
positions, they receive significantly higher benefits than every
other agent while making the same total investment. The system thus
evolves into a state where payoffs are unfairly~distributed.

The evolving network displays unfairness also in a second aspect. As
highly connected agents spend the same amount of resources as every
other agent, their contribution to any of their collaborations is
necessarily small. So collaborating with a highly connected agent
generally implies that one has to carry a large fraction of the
investment. Thus, the existence of highly connected agents implies
both unfairness in the global payoff distribution and unfairness in
the interaction-specific load distribution.

In the present paper, we investigate if a fairer load distribution
can be achieved if additional resources are available to agents of
high centrality. We extend the model class studied
in~\cite{DoRudolfGross} by including that an agent's success feeds
back on his cooperative investments. We show that the additional
feedback loop reduces the unfairness in the distribution of loads
but intensifies the unfairness in the distribution of~payoffs.

The paper is organized as follows: We start with a short summary of
the original model and outline the basic results. This will also
give us the opportunity to introduce the conventions needed. We then
include the additional feedback loop, discuss its effects on the
emergence of coordination and differentiation and study the
implications for fairness.

\section{Basic Model}
Consider a population of $N$ agents engaged in bilateral
interactions. The agents can for instance be people maintaining
social contacts, scientists collaborating on some project, or
companies entering business relationships. Every agent can invest
time/money/effort into each of the $N-1$ potential interactions with
another agent. Furthermore the $N^2-N$ individual amounts $e_{ij}$,
invested by agent $i$ into the interaction with agent $j$ can be
adapted selectively, independently, and continuously by the agents.
In other words, every agent is free to chose the amount of resources
invested into the collaboration with every single other agent.
Neither the total investment nor the structure of the collaboration
network are imposed a priori.

One can imagine that over time the population approaches an
equilibrium in which many potential interactions receive no
investment, while others are reinforced, forming links in a complex
network of cooperation. But, how will this network look like? How
will the investments be distributed? And will the network be fair in
the sense that all agents benefit in equal measure?

\begin{figure}[H]
\centering
\includegraphics[width=.85\textwidth]{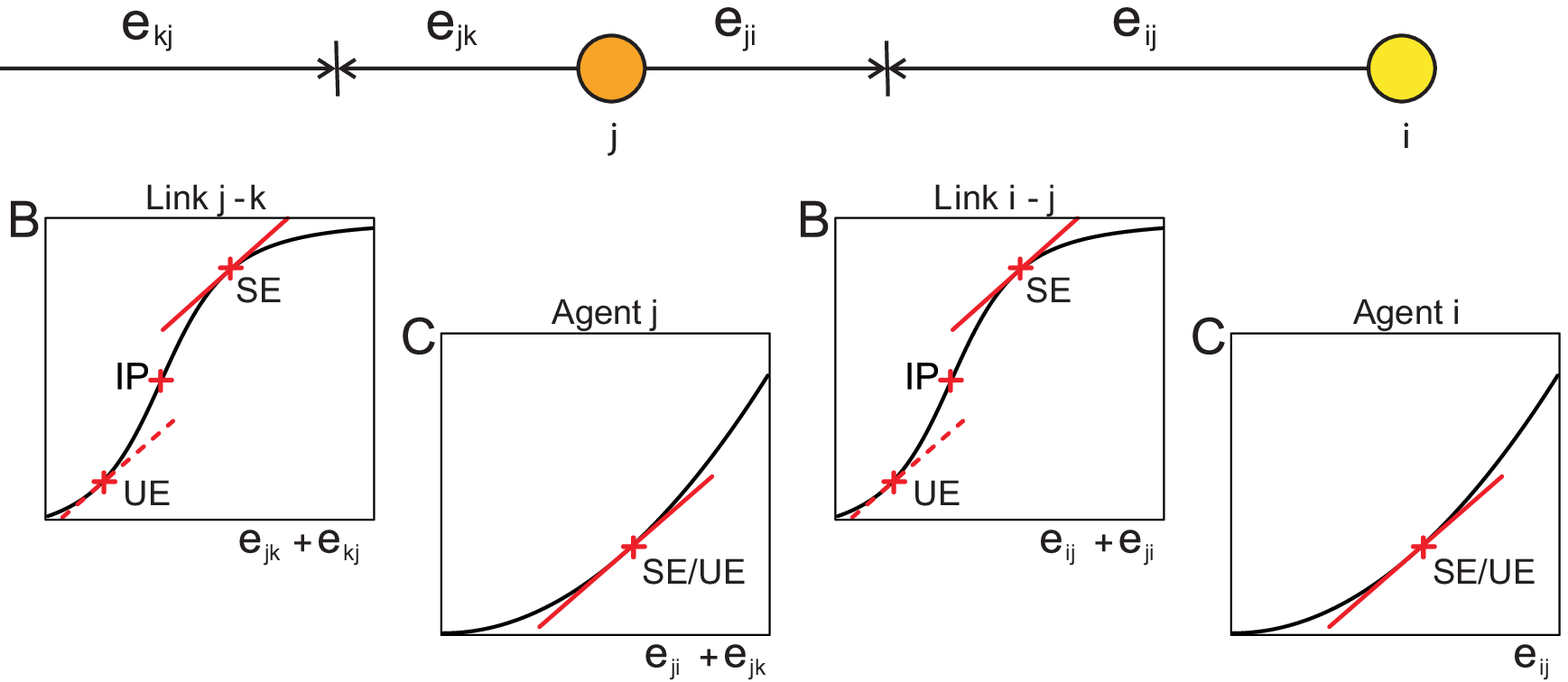}
\sffamily{\caption{(Reprinted from~\cite{DoRudolfGross}) Adjustment
of investments. Shown are the perceived cost functions $C$ and
benefit functions $B$ (insets) for the example of an agent i of
degree one interacting with an agent j of degree two (sketched). The
function $B$ depends on the sum of the agents investments into the
interaction while $C$ depends on the sum of all investments of the
agent. In every equilibrium (SE or UE) stationarity demands that the
slope of these functions be identical. This requires that the agents
make identical total investments. In stable equilibria (SE), the
operating point lies in general above the inflection point (IP) of
$B$, whereas equilibria found below the IP are in general unstable
(UE). Therefore, in a stable equilibrium both links produce the same
benefit and both agents make the same total investment. Iterating
this argument along a sequence of bidirectional links yields the
coordination properties (i) and (ii). } \label{pocfig:1}}
\end{figure}

Let us assume rational agents trying to maximize some payoff. A
generic model for a single interaction is the continuous snowdrift
game~\cite{DoebeliHauert}. In this game the payoff is $P=B-C$, where
$B$ and $C$ are non-linear functions. The benefit function $B$
depends on the sum of both investments while the cost function $C$
depends only on the investment of the agent under consideration.
While we do not restrict $B$ and $C$ to specific functional forms,
we assume that $B$ is sigmoidal and $C$ is superlinear (see
Figure~\ref{pocfig:1}). This captures basic features of real-world
systems such as inefficiency of small investments, saturation of
benefits, and additional costs incurred by overexertion of personal
resources.

In order to allow for multiple bilateral interactions per agent, let
us extend the snowdrift game by assuming that the benefits received
add linearly, while the cost is a function of the sum of investments
made by an agent. The payoff received by agent $i$ from the
interaction with an agent $j$ can then be written~as
\[P_{ij}=B\left(e_{ij}+e_{ji}\right)-\frac{e_{ij}}{\sum_{k}e_{ik}}\ C\left(\sum_{k}e_{ik}\right)
\] where we have allocated a proportional share of the total cost incurred by $i$ to the interaction with $j$.
We let the agents maximize their payoff dynamically in time by
following a steepest ascent approach~\cite{Snyman}
\begin{equation}\label{timeevolution}
        {\rm \frac{d}{dt}}e_{ij}= \frac{\partial}{\partial e_{ij}} \sum_{k} P_{ik} 
\end{equation}

By locally adapting their investments in the direction of the
steepest incline of payoff, agents may change the selection as well
as number of their collaborators, the amount invested in any
specific collaboration as well as the amount invested in total.

\subsection{Coordination of Investments}
In simulations the system shows frustrated, glass-like behavior;
starting from a homogeneous initial configuration, in which all
potential links are realized with identical investment plus a small
stochastic fluctuation, the system approaches either one of a large
number of different final configurations, which are local maxima of
the total payoff. To describe these configurations, the following
naming conventions are advantageous: Below, interactions which do
receive no investments such that $e_{ij}+e_{ji}=0$ will be denoted
as \emph{vanishing} interactions. Non-vanishing interactions will be
denoted as \emph{links}. Further, a set of agents and the links
connecting them are said to form a bidirectionally-connected
community (BCC) if every agent in the set can be reached from every
other agent in the set by following a sequence of bidirectional
(reciprocal) links.

In~\cite{DoRudolfGross}, it was shown analytically that all final
configurations share certain properties. Thus, within every evolved
BCC (i) every node makes the same total investment, and (ii) every
link produces the same benefit. The properties (i) and (ii) are
essential for a solution of the ODE system
Equation~\eqref{timeevolution} to be stationary and stable
(Figure~\ref{pocfig:1}). They thus apply to all stable steady
states.

\subsection{Differentiation of Payoffs}
The properties (i) and (ii) point to a remarkable degree of
coordination inside a BCC. This coordination results from the
selective and adaptive interaction of self-interested agents and is
achieved although no agent has sufficient information to estimate
the investment of any other agent in the
network~\cite{DoRudolfGross}. Interestingly, the emergent
coordination of investments does not necessarily imply that the
evolving networks are fair: Since all links in the BCC produce an
identical benefit, the total benefit received by an agent is
proportional to his degree, \textit{i.e.},~to the number of his
collaborations. Agents of high degree thus receive significantly
higher benefits while making the same investment as every
other~agent.

\begin{figure}[H]
\centering
\includegraphics[width=.5\textwidth]{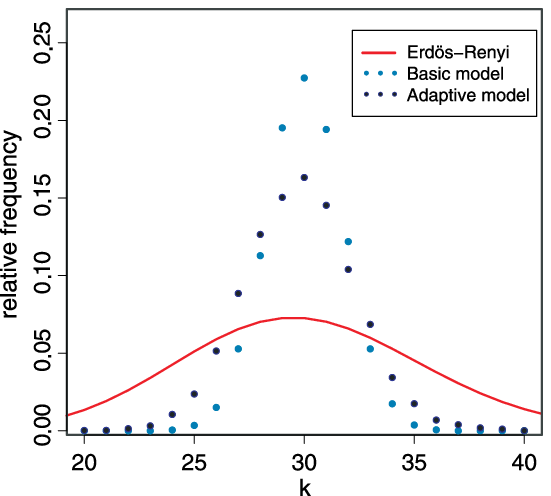}
\sffamily{\caption{Emergent heterogeneity in self-organized
networks. In comparison to a random graph (red), the degree
distribution of networks evolved in the basic model is relatively
narrow (light blue), but broadens as cost reduction for successful
agents is introduced (dark blue). All simulations rely on the
functions
$B=\,\tfrac{2\rho}{\sqrt{\tau+\rho^2}}+\tfrac{2(e_{ij}+e_{ji}-\rho)}{\sqrt{\tau+\left(e_{ij}+e_{ji}-\rho\right)^2}}$,
$C= \mu\left(\sum_{k}e_{ik}\right)^2$, $R=
1+\nu\,\sum_{k}B(e_{ij}+e_{ji})$. Parameters are chosen to obtain
networks with identical mean degree (basic model: $\rho=0.1$,
$\tau=0.124$, $\mu=2.731$, adaptive model $\rho=0.395$, $\tau=0.1$,
$\mu=2.32$, $\nu=0.05$). Results are averaged over 1000 networks of
size $N=100$. \label{pocfig:2}}}
\end{figure}

Figure~\ref{pocfig:2} shows a representative degree distribution
$p_k$ specifying the relative frequency of nodes with degree $k$ of
an evolved network in the final state. Although agents follow
identical rules and the network of collaborations is initially
almost homogeneous, the distribution has a finite width indicating a
certain heterogeneity. However, the distribution is narrower, and
therefore fairer, than that of an Erd\"os--R\'enyi random graph,
which constitutes a null-model for network topology.

The emergence of the payoff disparity can be traced to the discrete
nature of the links. As we reported above the local optimization
carried out by the agents leads both to a coordination of
investments in the links and to a local optimization of the total
payoff. For a given choice of parameters, the optimal payoff will be
extracted if a certain number of collaborations exist in average per
agent. However, the optimal total number of links is not necessarily
commensurable with the number of agents and hence the maximal total
payoff can only be extracted when the collaborations are distributed
unfairly.

\subsection{Imbalance in Load Distributions}
In order to sustain the extraction of high payoffs by agents with
high degree, investments have to be redistributed across the
network. While the transport of resources is not explicitly included
in the model, it enters through the asymmetry of the individual
interactions. Consider for instance an agent of degree one. This
agent has to focus his investment on a single link. The partner
participating in this link will therefore only need to make a small
investment in the interaction to make it profitable. He is thus free
to invest a large portion of his total investment into links to
other agents of possibly even higher degree. In this way,
investments flow toward regions of high connectivity where large
payoffs are extracted.

The most extreme case for an unequal load distribution within a
cooperation is realized in unidirectional links. These correspond to
interactions, in which one partners invests without any
reciprocation. While the behavior of the exploited agent seems
irrational, the analysis in~\cite{DoRudolfGross} shows that it can
arise in a population of rational self-interested agents.
Simulations reveal that unidirectional investments are not even
rare: Depending on the mean degree of the evolving network, up to
50\% of all cooperative links can be
unidirectional~\cite{DoRudolfGross}.

\section{Adaptive Model}
As shown above, the individual selection of eligible cooperation
partners promotes the coordination of cooperative investments, the
differentiation of received payoffs but also the emergence of
unequal workloads within a cooperation. One can now argue that in
the real world successful agents have access to more resources,
which could allow them to reciprocate more strongly in their
collaborations, which would in turn lead to a fairer load sharing.
Below, we study the effect of an agent's success feeding back on his
cooperative investments. Including a benefit-dependent reduction of
$C$ in the model yields a fully adaptive
network~\cite{GrossBlasius}.

In our adaptive model, agents enjoy benefit-dependent cost reduction
    \begin{equation}
    P_{ij}=B\left(\sigma_{ji}\right)-\frac{e_{ij}}{\Sigma_{i}}\ C\left(\Sigma_i\right)\cdot \frac{1}{R\left(\beta_i\right)}
\end{equation}
where $\sigma_{ji}\!:=e_{ij}+e_{ji}$, $\Sigma_i\!:=\sum_{k}e_{ik}$,
and $R$ is a monotonically increasing function of the total benefit
of agent $i$
\mbox{$\beta_i\!:=\!\sum_{k}B\left(\sigma_{ik}\right)$}. As above,
we assume the benefit function $B$ to be sigmoidal. Moreover, we
assume the cost function to be super-linear and of the general form
$C\left(\Sigma_i\right)\propto\left(\Sigma_i\right)^\gamma$.

\begin{figure}[H]
\centering
\includegraphics[width=9cm]{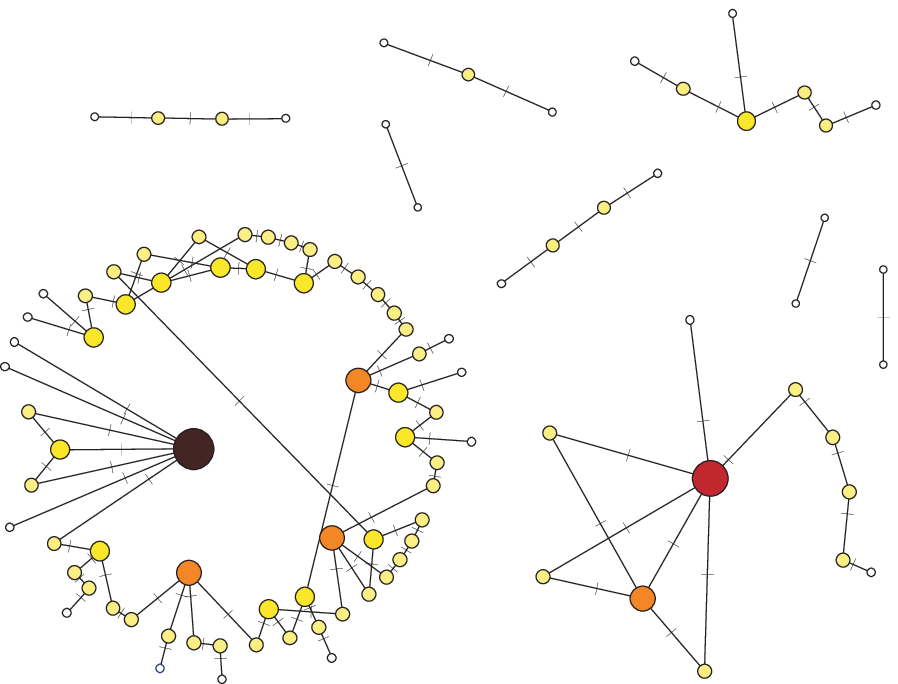}
\sffamily{\caption{Self-organized network evolved in the adaptive
model. Nodes represent agents, while each link represents a
non-vanishing cooperative interaction. The small dash on the link is
a fairness indicator: the further it is shifted toward one agent,
the lower is the fraction of the total investment into the
cooperation that he contributes. Nodes extracting more payoff are
shown in darker color and are placed toward the center of the
community. The size of a node indicates the total investment the
agent makes. In the final configuration all links within a BCC
receive the same total investment and all nodes of the same degree
make the same total investment. Simulation parameters: $\rho=0.7$,
$\tau=0.1$, $\mu=2.24$, $\nu=0.588$} \label{pocfig:3}}
\end{figure}

\enlargethispage*{1cm} Below, we show that in the adaptive model
property (ii) still holds while property (i) needs to be modified:
The total amount of investment differs among agents within a BCC as
agents enjoy \mbox{benefit-dependent} cost reduction
(Figure~\ref{pocfig:3}). However, we find that agents of the same
degree approach the same investment level. Consequently distinct
classes of agents arise which differ both in investment and
in~payoff.

\subsection{Coordination and Differentiation}

For deriving the modified coordination properties (i) and (ii) we
proceed analogously to the \mbox{non-adaptive} case,
\textit{i.e.},~we evaluate the conditions for a solution of the ODE
system Equation~\eqref{timeevolution} to be stationary and stable.

First the stationarity condition. Defining
$\partial_x:=\frac{\partial}{\partial x}$, we can rewrite the
stationarity condition
\begin{equation}
    {\rm \frac{d}{dt}}e_{ij}=0= \partial_{e_{ij}}\left[ \sum_{k} B\!\left(\sigma_{ik}\right)-\frac{C\!\left(\Sigma_{i}\right)}{R\left(\beta_{i}\right)}\right]
\end{equation}
as
\begin{align}
\begin{split}
\partial_{e_{ij}}B\!\left(\sigma_{ij}\right)&=\frac{\partial_{e_{ij}}C\!\left(\Sigma_{i}\right)}{R\left(\beta_{i}\right)}-\frac{C\!\left(\Sigma_{i}\right)}{\left(R\left(\beta_{i}\right)\right)^2}\,\partial_{\beta}R\left(\beta_{i}\right)\,\partial_{e_{ij}}\beta    \\
&=\frac{R\left(\beta_{i}\right)}{\left(R\left(\beta_{i}\right)\right)^2+\partial_{\beta}R\left(\beta_{i}\right)}\:\frac{\partial_{e_{ij}}C\!\left(\Sigma_{i}\right)}{C\!\left(\Sigma_{i}\right)}
\end{split} \label{stat-adapt_eq}
\end{align}
where we used
$\partial_{e_{ij}}\beta=\partial_{e_{ij}}B\!\left(\sigma_{ij}\right)$.
The right hand side of Equation~\eqref{stat-adapt_eq} does only
depend on the node parameters $\Sigma_i$ and $\beta_i$,
\textit{i.e.},~in a steady state
$\partial_{e_{ij}}B\left(\sigma_{ij}\right)$ is identical for all
bilateral links $ij$ of agent $i$. From
$\partial_{e_{ij}}B\!\left(\sigma_{ij}\right)=\partial_{e_{ji}}B\!\left(\sigma_{ij}\right)$
it then follows that all bilateral links of $j$, and, by iteration,
that all bilateral links within one BCC are identical with respect
to $\partial_{e_{mn}}B\left(\sigma_{mn}\right)$. Since the benefit
function is sigmoidal, a given slope can be found in at most two
points along the curve: one above and one below the inflection point
(IP) (Figure~\ref{pocfig:1}). This implies that if a stationary
level of investment is observed in one link, then the investment of
all other links of the same BCC is restricted to one of two
operating~points.

In the basic model, stability analysis revealed that the operating
point below the IP is unstable and can thus be ruled out
(Figure~\ref{pocfig:1}). Unfortunately, in the extended model,
the analysis cannot be performed to the same extent. However, in
extensive numerical simulations we have not observed a single
equilibrium which contained a link operating below the IP. This
strongly indicates that the dynamics of the extended model are
governed by similar stability conditions as the dynamics of the
basic model, which reproduces property (ii):
\begin{equation}\label{prop2}
    \sigma_{ij}\equiv\sigma \ \forall \ \text{links} \ ij \ \text{in a BCC}
\end{equation}

Combining Equations~\eqref{stat-adapt_eq} and~\eqref{prop2}, we can
now derive property (i): Let us consider a single BCC. According to
Equation~\eqref{prop2}, the total benefit of an agent $i$ in this
BCC is a function of its degree $d_i$:
    \[\beta_i=\sum_{k}B\left(\sigma_{ik}\right)=d_i\cdot B\left(\sigma\right)
\]
Inserting this relation in Equation~\eqref{stat-adapt_eq}, we find
that the left hand side is constant, while the first factor on the
right hand side only depends on $d_i$. The second factor on the
right hand side is injective, as we assumed
$C\left(\Sigma_i\right)\propto\left(\Sigma_i\right)^\gamma$. It thus
follows that nodes of the same degree have to make the same total
investment $\Sigma_i$, even if they are only connected through a
chain of nodes making different investments. However, nodes of
different degree $d_i$ can differ in their total investment.

The emergence of distinct classes of nodes, which differ in degree
(and therefore in payoff) and total investment, is illustrated in
Figure~\ref{pocfig:3}. The figure shows the final configuration of
an exemplary model realization with 100 nodes. Nodes of high degree
received high payoffs (coded in the node color) and run high total
investments (coded in the node size).

Compared to the basic model, the adaptive model leads to
considerably broadened degree distributions (Figure~\ref{pocfig:2}). We can thus conclude that the additional
resources available to high degree agents are at least in part used
to establish additional links. This leads to an increased income
disparity in the evolving~network.

\subsection{Fairness of Load Distribution}
Let us now address the fairness of individual interactions. As in
the basic model, also in the adaptive model the investments of two
interacting agents into a common collaboration are usually
asymmetric. In Figure~\ref{pocfig:3}, this is apparent in the
position of the fairness indicators on the links: The further it is
shifted toward one agent $i$, the lower the fraction
$e_{ij}/\sigma_{ij}$ that he contributes. Even in small network
components, the fairness indicators reveal a flow of investments
towards regions of high connectivity; as a general rule high-degree
nodes contribute less to an interaction than their lower-degree
partner.

In both the basic and the adaptive model, the specific load
distribution in an interaction depends on the exact topological
configuration of the respective network component. Hence, for
comparing the fairness of load distributions in both models, it is
necessary to consider components of the same structure.

The simplest degree-heterogeneous structure is a chain of three
nodes $i,j$ and $k$. In such a structure, the two degree-one nodes
$i$ and $k$ necessarily concentrate all their investment in the
cooperation with the middle node $j$, while the latter splits its
investment in equal parts. The fraction $e_{xj}/\sigma_{xj}$ that
the middle node contributes to each of the two links can be
calculated as
    \[ \frac{e_{xj}}{\sigma_{xj}}=\frac{0.5\Sigma_j}{0.5(\Sigma_i+\Sigma_j+\Sigma_k)}, \qquad x=i,j 
\]

In the basic model, the total investment of all three nodes are
identical. Thus,  $e_{xj}/\sigma_{xj}=1/3$. In the adaptive model,
$\Sigma_j>\Sigma_i=\Sigma_k$. Thus, $e_{ij}/\sigma_{ij}>1/3$,
\textit{i.e.},~the load distribution is fairer than in the basic
model (cf. fairness indicators on three node chain in
Figure~\ref{pocfig:3}).

Generalizing the reasoning sketched above, we find that for any
given topological configuration, the imbalance in the load
distribution is milder in the adaptive model than in the basic
model. We can thus conclude that the additional resources available
to high degree agents are partly reinvested in existing links
enhancing the fairness of the respective interactions.

Further confirmation for fairer load distributions in the adaptive
model comes from the numerical data: In extensive simulations using
a wide range of parameters we have not observed a single
unidirectional link. This observation stands in sharp contrast to
the observations made in the basic model, where unidirectional
links---the most extreme case of unequal load
distribution---constitute a considerable fraction of all links in a
network.

\section{Summary}
In this paper, we have extended a recently studied model for the
formation of cooperation networks by taking into account that an
agent's success feeds back on his cooperative investments. Although
agents have large freedom in their investment strategy and little
information about investments of others, we find that like the basic 
model, the adaptive self-organizes toward configurations exhibiting a
high degree of coordination: In all final configurations,
bidirectionally connected communities approach a state in which the
benefit produced by each link is identical and in which the total
investment made by a agent is either identical (basic model) or
falls into distinct classes (adaptive model).

Despite coordination, both models display unfairness in two aspects:
payoffs are unequally distributed in the population and loads are
unequally distributed between cooperating partners. Both aspects can
be traced back to the local payoff optimization governing the
dynamics of the system. The optimization of payoffs implies an
optimal number of links in the system. The latter, however, is
usually incommensurable with the number of agents leading to
configurations, in which some agents have more links,
\textit{i.e.},~a higher degree, than others. In both versions of the
model, agents of higher degree are found to extract more payoff and
contribute less to a cooperation than their lower degree partners.

In the adaptive model, cost reduction for successful agents makes
additional resources available to highly-connected agents. These
resources are partly invested in existing collaborations, leading to
fairer load distributions, but also in establishing new
collaborations, leading to broadened degree distributions.

Let us emphasize that  differentiation and emergence of unfairness
in an initially homogeneous population has previously been discussed
in the context of the adaptive
networks~\cite{ItoKaneko,Zimmermann2000,EguiluzZimmermannCela}.
However, to our knowledge the here proposed framework is the first,
in which the phenomena can be linked to details of the dynamic
self-organization process. Our analysis has greatly profited from
the dual nature of the model class under consideration. The
continuous nature allowed for a model description in terms of
ordinary differential equations and thus for an analysis with the
tools of dynamical systems theory. On the other hand, the discrete,
unweighted nature of the final configurations allowed us to use the
concepts of graph theory.

\bibliographystyle{mdpi}
\makeatletter
\renewcommand\@biblabel[1]{#1. }
\makeatother


\end{document}